\documentclass{article}

\def\isTechReport{1}
\if\isTechReport1
\newcommand{\techReportVersion}[1]{#1}
\newcommand{\conferenceVersion}[1]{}
\fi
\if\isTechReport0
\newcommand{\techReportVersion}[1]{}
\newcommand{\conferenceVersion}[1]{#1}
\fi

\usepackage{enumitem}
\usepackage{amsmath}
\usepackage{url}
\usepackage{graphicx}

\newcommand{\str}[1]{\text{``}\texttt{#1}\text{''}} 

\begin{document}

\title{Strategies for basing the CS theory course on non-decision
  problems}

\author{
\hspace{2mm}John MacCormick\footnote{Also affiliated with School of Computing Sciences, University of East Anglia, 2017--19}\\ [1mm]
Dickinson College\\
}
\date{24 November 2017}

\maketitle
\begin{abstract}
  \techReportVersion{\emph{This is an extended
      version of a paper published in the proceedings of SIGCSE 2018.} \par} Computational and complexity theory
  are core components of the computer science curriculum, and in the
  vast majority of cases are taught using \emph{decision problems} as
  the main paradigm.  For experienced practitioners, decision problems
  are the best tool. But for undergraduates encountering the material
  for the first time, we present evidence that \emph{non-decision
    problems} (such as optimization problems and search problems) are
  preferable. In addition, we describe technical definitions and
  pedagogical strategies that have been used successfully for teaching
  the theory course using non-decision problems as the central
  concept.
\end{abstract}

\section{Introduction and related work}
\label{sec:intro}

Many undergraduate computer science curricula include a ``theory''
course, covering some aspects of automata theory, computability, and
complexity.  Central questions addressed by a theory course include:
(a)~Given a computational problem $C$, can we solve $C$ using a
computer program? (b) If so, can we do so in polynomial time?  A
strong majority of popular theory textbooks focus on computational
problems that are in fact \emph{decision problems}---problems that
have a yes/no answer.  An alternative approach, advocated by this
paper, is to consider \emph{general} computational
problems---including non-decision problems such as \emph{optimization
  problems} and \emph{search problems}---while narrowing in on
decision problems when appropriate.  

There are two principal objectives of this paper. First, we explain
why non-decision problems are pedagogically advantageous for
introductory theory courses, supporting the explanation with empirical
evidence based on a survey of computer science majors. Second, we
describe technical details and pedagogical strategies for using
non-decision problems as the central concept in the CS theory course,
based on experience with four years of course offerings and a
recent textbook~\cite{MacCormick2018PUP}.  The use of non-decision
problems is the single most important idea for the suggested approach,
but additional strategies for making the CS theory course accessible
are also described. This includes the use of real computer programs
processing ASCII strings, instead of the more common model of Turing
machines processing symbols from an arbitrary
alphabet. \techReportVersion{Ultimately, this approach could be used
  to teach the theory course earlier in the curriculum and to a more
  diverse audience. }

The nature of the audience is crucial to our considerations. In this
paper, the primary target is undergraduate students who are seeing
computability and complexity theory for the first, and quite possibly
the last, time. We will refer to such students as the
\emph{novice audience}. We focus only on the novice audience in this
paper, ignoring other theory course scenarios such as graduate
courses or hybrid courses taught to a mixture of graduate students
and advanced undergraduates.  We also ignore the possibility of theory
courses that focus mostly on automata theory, assuming instead that
computability and complexity theory comprise at least a majority of
the course. 

For this audience, the key advantage of non-decision problems is that
they are more realistic: they match the previous programming and
algorithms experience of undergraduates more closely. As a concrete
example (details are given later), a decision problem may ask the
yes/no question, ``Does this graph have a Hamilton cycle?'' The
corresponding non-decision problem is, ``Please give me a Hamilton
cycle of this graph if it has one.'' A program that solves the
non-decision version produces a useful result that could conceivably
be used in a real-world application, whereas a program solving the
decision version yields only a single bit: it tells us whether or not
a Hamilton cycle exists, but gives no additional information. For the
novice audience, focusing on these single-bit decision problems
potentially positions the course as abstruse and irrelevant. Focusing
on non-decision problems with meaningful solutions, on the other hand,
provides more direct connections to earlier courses. In
section~\ref{sec:empirical}, we provide a combination of educational
theory and empirical evidence to support these claims.

However, it is important to note at the outset that we do not suggest
decision problems should be jettisoned from the theory course. On the
contrary, there are some compelling reasons to incorporate decision
problems into any theory course, including: (i)~some theorems and
proofs are more concise and elegant when phrased in terms of decision
problems; (ii)~the vast majority of existing CS theory literature
considers only decision problems, so it is essential that any student
planning to take a subsequent graduate-level theory course has been
exposed to the decision problem framework. Therefore, in this paper we
advocate a hybrid approach, whereby the earlier topics in the theory
course are taught with a focus on non-decision problems.  When
NP-completeness is introduced---typically sometime in the second half
of a one-semester course---we suggest transitioning to the classical
approach and focusing only on decision problems (with some minor
exceptions described later). This hybrid approach allows the novice
audience to establish an initial appreciation of the relevance and
importance of CS theory via non-decision problems, while still being
well-versed in the classical approach by the end of the course.

\techReportVersion{
\begin{figure}
  \centering
  \begin{tabular}{l}
Arora, Barak (2009)~\cite{Arora2009}\\
Davis, Sigal, Weyuker (1994)~\cite{davis1994computability}\\
Goldreich (2010)~\cite{goldreich2010p}\\
Hopcroft, Motwani, Ullman (2006)~\cite{hopcroft2006introduction}\\
Lewis, Papadimitriou (1997)~\cite{lewis1997elements}\\
Linz (2011)~\cite{linz2011introduction}\\
Moore, Mertens (2011)~\cite{Moore2011}\\
Papadimitriou (1994)~\cite{Papadimitriou1994}\\
Rich (2007)~\cite{rich2007automata}\\
Sipser (2013)~\cite{Sipser2013}\\
  \end{tabular}
  \caption{Selection of CS theory textbooks used for comparison.}
  \label{fig:textbooks}
\end{figure}
}

As a basis for comparing and contrasting various approaches to the
theory course, we will refer to a selection of ten textbooks covering
the relevant
material~\cite{Arora2009,davis1994computability,goldreich2010p,hopcroft2006introduction,lewis1997elements,linz2011introduction,Moore2011,Papadimitriou1994,rich2007automata,Sipser2013}\conferenceVersion{;
  please see the technical
  report~\cite{MacCormick:NonDecProbs:techreport:2017} for details
  about the selection of comparison textbooks}.
\techReportVersion{These books are listed in
  Figure~\ref{fig:textbooks}. The selection emphasizes diversity and
  there has been no attempt to include all relevant books. Of our ten
  selected books, the following probably come closest to being standard
  CS theory treatments: Hopcroft \emph{et al.}, Linz, Rich, and
  Sipser. Brief descriptions of the remaining six books are as
  follows: Arora/Barack and Moore/Mertens are significantly more
  advanced; Davis \emph{et al.}, Lewis/Papadimitriou, and Papadimitriou
  are classics; Goldreich is nonstandard and advocates the use of
  nondecision problems as in this paper.}

\section{Related work}
\label{sec:related-work}




There are several strands of related work aiming to make CS theory
courses more practical, accessible, or meaningful.  We might classify
these strands as: (i)~interactive automata software tools such as
JFLAP and
DEM~\cite{Chesnevar2003,rodger2006jflap,rodger2009increasing};
(ii)~lab assignments and visualizations for NP-completeness, an
approach sometimes described as ``NP-completeness for
all''~\cite{Crescenzi2013,enstrom2010computer,Lobo2006};
(iii)~recasting the theoretical ideas themselves, for example by
emphasizing non-decision problems and including holistic discussions
about the implications of NP-completeness and
P-versus-NP~\cite{fortnow2013golden,Goldreich2006,goldreich2010p,mandrioli1982teaching,papadimitriou1997np}.
The present paper falls firmly in category (iii). The ideas described
here are orthogonal to categories (i) and (ii). Indeed, the ideas of
this paper have been employed over a four-year period in a course that
also uses JFLAP and practical programming assignments, thus
benefiting from all three strands of literature on the CS theory
course.

To the best of our knowledge, Goldreich's position
paper~\cite{Goldreich2006} and subsequent
textbook~\cite{goldreich2010p} comprise the previous work most similar
to the present paper.  Goldreich makes several important, insightful,
and useful suggestions for improving the theory course, including
strong advocacy for the use of non-decision problems.
The novel contribution of the present paper is to recast the use of
non-decision problems in a manner that is more accessible to the
novice audience, thus enabling instructors to focus on non-decision
problems and deliver the consequent educational benefits to
students.

\section{A practical definition of computational problems}
\label{sec:practical-definition}

Before the advantages of non-decision problems can be explained, we
review some elementary background material to establish notation and
give a formal definition of a ``computational problem,'' which can of
course be either a decision problem or a non-decision problem. The
background review in this section also includes several ancillary
recommendations for how to present this material to a novice
audience.

Computational theory is often developed in terms of an arbitrary
\emph{alphabet} (i.e.\ a finite set of symbols) denoted $\Sigma$. The
set of all possible \emph{strings} (i.e.\ finite sequences of symbols)
on $\Sigma$ is denoted $\Sigma^*$.
Experienced practitioners understand that the choice of alphabet is
irrelevant for most purposes.  But for students being introduced to
complexity theory for the first time, it may be preferable to employ a
more familiar and obviously relevant alphabet---one that is used by
programmers to describe the inputs and outputs of computer programs in
practice. This motivates us here to take the ASCII alphabet as our
primary example of~$\Sigma$.

For similar reasons, we use real computer programs as our main
computational model, rather than the Turing machines which are more
common in the theory literature. A program $P$ receives a single ASCII
string $w$ as input and the output is either undefined (e.g.\ if $P$
crashes or enters an infinite loop) or is an ASCII string denoted
$P(w)$.  When $P(w)$ is defined, we say $P$ \emph{rejects} $w$ if
$P(w)=\str{no}$ and $P$ \emph{accepts} $w$ if $P(w)\neq\str{no}$.

A central concept in complexity theory is a \emph{language} (sometimes
called a \emph{formal language}), defined as a subset of $\Sigma^*$.
The intuitive notion of ``solving a problem'' is usually formalized as
``deciding a language.''  Thus, a central concept in many treatments
is that a program $P$ \emph{decides} a language $L$ if $P$ accepts all
strings $s\in L$ and rejects all $s\notin L$.

What is the connection between ``deciding a language'' and ``solving a
problem''?  For so-called \emph{decision problems}, there is a direct
and simple connection.  Informally, decision problems are
computational questions that have a yes/no answer, such as ``Given an
integer $m>0$, is $m$ prime?'' or ``Given a graph $G$, does $G$
contain a Hamilton cycle?''.  Formally, one could define a decision
problem as a function from $\Sigma^*$ to $\{\str{yes},\str{no}\}$.
When taking $\Sigma$ as the ASCII alphabet, we first agree an encoding
of the relevant mathematical objects into ASCII. For example: the
integer $m=43552$ might be encoded as the ASCII string \str{43552}; a
graph that is a three-cycle with three vertices labeled $a,b,c$ might
be encoded as the ASCII string \str{a,b b,c c,a}.

For a given decision problem $D$, strings for which the answer is
\str{yes} are called \emph{positive instances} of $D$ and all other
strings are \emph{negative instances}.  This leads to a simple and
obvious mapping between languages and decision problems. Any language
$L$ corresponds to a decision problem $D_L$ defined as follows: for
$s\in \Sigma^*$, $D_L(s)=\str{yes}$ if and only if $s\in L$. The
reverse mapping is also simple and obvious: a decision problem $D$
corresponds to a language $L_D$ with the property that $s\in L$ if and
only if $D_L(s)=\str{yes}$.

As a result of the exact correspondence between decision problems and
languages, the theory of computability and complexity is usually
described in terms of languages.  As already mentioned in the
introduction, the advantages of doing so include simple notation and
compact statements of certain theorems, but disadvantages for the
novice audience include a high level of abstraction and the
restriction to decision problems, which may appear unfamiliar and
irrelevant.


To illustrate the distinction between decision and non-decision
problems, we will use two running examples throughout the paper:
\begin{itemize}[leftmargin=*]
\item \textsc{Factor}: The input is a string representing a positive
  integer $m$ in decimal notation (e.g.\ \str{35}), and a solution is
  any factor of the input other than 1 and $m$ (e.g.\ \str{5} or
  \str{7} for $m=35$), or \str{no} if no such factor exists.
  \textsc{FactorD} is the corresponding decision problem, with
  solution \str{yes} if $m$ has a non-trivial factor and \str{no}
  otherwise.
\item \textsc{HamCycle}: The input is a string representing a graph
  $G$ (e.g. \str{a,b b,c c,a}), and a solution is any Hamilton cycle
  of $G$ (e.g.\ \str{a,b,c} for the given example of $G$), or \str{no}
  if no such cycle exists. \textsc{HamCycleD} is the corresponding
  decision problem, with solution \str{yes} if $G$ has a Hamilton
  cycle and \str{no} otherwise.
\end{itemize}
We also assume the reader is familiar with \textsc{Sat} (which asks
for a satisfying assignment to a Boolean formula) and \textsc{SatD}
(the decision variant, which asks whether a satisfying assignment
exists).


Several important differences between the decision and non-decision
variants should be immediately obvious from the \textsc{Factor} and
\textsc{HamCycle} examples:
\begin{itemize}[leftmargin=*]
\item Non-decision problems can have multiple solutions (e.g.\ an
  integer $m$ can have multiple non-trivial factors; a graph $G$ can
  have multiple distinct Hamilton cycles), whereas decision problems
  always have a unique solution (\str{yes} or \str{no}).
\item It is possible for decision and non-decision variants of the
  same problem to have apparently different complexity
  properties. \textsc{Factor} is a classic example here, since
  \textsc{FactorD} can be solved in polynomial time by the AKS
  algorithm~\cite{Agrawal2004}, whereas \textsc{Factor} has no known
  polynomial-time method of solution.\footnote{For NP-complete
    problems like \textsc{HamCycleD}, however, this never happens:
    given a non-decision problem $C$ whose decision variant is an
    NP-complete problem $C'$, it is known that $C$ always has a
    polynomial time reduction to $C'$. This property is sometimes
    known as \emph{self-reducibility}~\cite{goldreich2010p}.}
\item Non-decision variants appear to be more ``natural.''  For
  example, it is hard for the novice audience to imagine an
  application where it is useful to determine the mere existence of a
  Hamilton cycle, rather than determining the sequence of vertices in
  the cycle.\footnote{In contrast, expert practitioners know that
    sometimes the single-bit decision \emph{is} useful. For example,
    determining whether an integer is prime or composite is important
    in cryptographic applications.}
\end{itemize}

A significant fraction of theory textbooks give no formal definition
of a non-decision problem. In our sample of ten books, seven give no
formal
definition~\cite{Arora2009,davis1994computability,hopcroft2006introduction,lewis1997elements,linz2011introduction,Moore2011,Sipser2013}. One~\cite{rich2007automata}
gives a brief definition; Papadimitriou~\cite{Papadimitriou1994} and
Goldreich~\cite{goldreich2010p} give detailed definitions and
analysis. These last two define non-decision problems in terms of
binary relations with certain technical properties: they are
\emph{polynomially decidable} and \emph{polynomially bounded} (or
\emph{balanced}). This approach is correct and concise, but perhaps
more abstract than necessary for a novice undergraduate audience. Instead, the following
concrete definition is recommended for the novice audience:
\begin{quote}
  A \emph{computational problem} (which may or may not be a decision
  problem) is a function $F$, mapping ASCII strings to sets of ASCII
  strings. If $F(x)=\{s_1, s_2, \ldots\}$, we call
  $\{s_1, s_2, \ldots\}$ the \emph{solution set} for $x$, and each
  $s_i$ is a \emph{solution} for $x$. If $F(x)=\{\str{no}\}$, then $x$
  is a \emph{negative instance} of $F$; otherwise $x$ is a
  \emph{positive instance}.  If $F$ has unique solutions, $F$ is a
  \emph{function problem}. For function problems, we can drop the set
  notation, for example writing $F(x)=y$ instead of $F(x)=\{y\}$.
\end{quote}
For example: \str{35} is a positive instance of \textsc{Factor}, and
we have $\textsc{Factor}(\str{35})=\{\str{5}, \str{7}\}$; \str{29} is
a negative instance, and we have
$\textsc{Factor}(\str{29})=\{\str{no}\}$. Note that neither
\textsc{Factor} nor \textsc{HamCycle} is a function problem, since
both problems can have multiple solutions.  The definition of
computational problem can obviously be adapted to non-ASCII alphabets,
but we omit those details here.

The above definition has an intuitive connection to computer programs
that produce meaningful output (as opposed to a single accept/reject
bit).  Formally, we say a computer program $P$ \emph{solves} the
computational problem $F$ if $P(x)\in F(x)$ for all $x$. That is, the
program always terminates and outputs a correct solution.

\section{Empirical evidence for pedagogical benefit of non-decision problems}
\label{sec:empirical}

The core motivation for our approach is the well-established
pedagogical principle that students learn new concepts more
effectively when those concepts are placed in a familiar context. The
effectiveness of learning is further enhanced if the new concept is
perceived as \emph{useful} or
\emph{applicable}~\cite{fink2013creating}. Non-decision problems
conform to these two criteria---familiarity and applicability---much
more closely than decision problems. The familiarity criterion is
indisputable, since almost every computer program written by students,
beginning with the first programming course and continuing throughout
their careers, computes meaningful answers to general problems rather
than producing a single-bit accept/reject decision. The second
criterion of perceived usefulness or applicability also seems
plausible, but requires supporting evidence. To investigate this, a
sample of 41 computer science majors were given descriptions of four
computer programs which solved decision and non-decision variants of
two different problems. Participants rated the usefulness of each
program on a Likert scale from 1 (extremely useful) to 5 (not at all
useful).
The order of presentation (decision versus non-decision variants) was
varied to eliminate ordering effects. We control for the intrinsic
perceived usefulness of any given problem by presenting decision and
non-decision variants of the \emph{same} problems. A total of $n=81$
paired (i.e. decision versus non-decision) responses were received
from the 41~participants.  \conferenceVersion{Further details of the
  survey design, demographics, and results are available in an
  accompanying technical report~\cite{MacCormick:NonDecProbs:techreport:2017}.}

\techReportVersion{Approximately 50\% of participants were sophomores,
  with the remainder distributed among first, third, and fourth-year
  students. All had completed either one or two introductory Java
  programming courses, and 85\% had completed a data structures
  course. On average, participants had completed six computer science
  courses at the time of taking the survey. For our purposes, this
  demographic is representative of the ``novice audience.''}

\begin{figure}
  \centering
  \includegraphics[width=60mm]{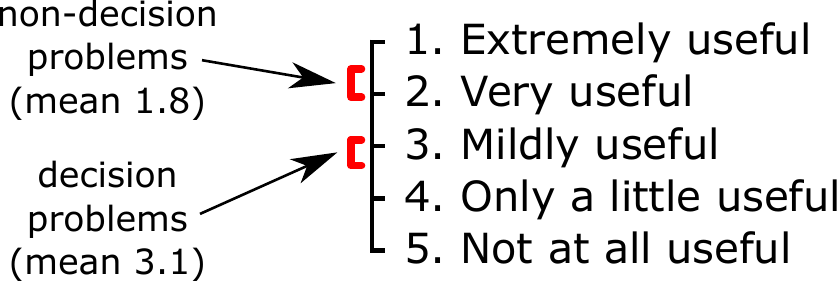}
  \caption{Students perceive programs solving non-decision problems as
    considerably more useful than programs solving decision
    problems. The red bars represent 99\% confidence intervals for the mean.}
  \label{fig:likert}
\end{figure}

The results show that programs solving non-decision problems are
perceived as significantly more useful than programs solving decision
problems (mean 1.8 vs 3.1 on the Likert scale, with standard
deviations of about 0.7 and 0.8 respectively, leading to standard
errors of less than 0.1 in both cases). A 99\% confidence interval
for the population mean, formed from plus-or-minus three standard errors, is shown on
figure~\ref{fig:likert}; the intervals suggest a large and significant
difference. Two more rigorous tests confirm this:
(i)~a Wilcoxon signed-rank test for asymmetry of the paired
differences has a negligible $p$-value ($p<10^{-11}$); (ii)~the same
test run on shifted data, with the non-decision responses shifted away
from ``useful'' by an entire Likert gradation, also has a negligible
$p$-value ($p<10^{-6}$).
We conclude that the difference in student perceptions of usefulness
is rather large---certainly more than one Likert gradation and hence
exceeding the distinction between ``very useful'' and ``mildly
useful,'' as shown on figure~\ref{fig:likert}.

\techReportVersion{The Wilcoxon signed-rank test is appropriate here
  because the responses are on an ordinal scale in which the numerical
  values themselves (1--5) have no meaning beyond their
  ordering. Nevertheless, if we are willing to assume an approximately
  normal distribution of the numerical responses we can perform the
  more familiar paired $t$-test for difference in means. This yields
  $p<10^{-19}$ on the raw responses and $p\approx 0.003$ on the
  shifted responses, leading to the same conclusions as the Wilcoxon
  tests.

  For simplicity, the Wilcoxon tests described above employed one
  assumption that is perhaps a little dubious: the responses of the
  same participant for different computational problems were treated
  as independent. However, additional tests in which the computational
  problems were analyzed separately also yield negligible $p$-values
  ($p<10^{-6}$ on both problems).}

It is worth emphasizing here that this difference in perceived
usefulness exists in a \emph{novice audience} (as defined in
section~\ref{sec:intro}), and this underpins the key pedagogical point
of the paper.  Experienced practitioners know that decision programs
can often be converted to equivalent non-decision programs
with only a logarithmic increase in running time. But because the
novice audience does not share this intuition, we should exploit the
novices' perception that non-decision programs are useful. We conclude
that elementary concepts in an introductory computer science theory course
should be taught using non-decision problems whenever possible.

\section{Computability for non-decision problems}
\label{sec:comp-non-decis}

The previous section established the key pedagogical advantages of
non-decision problems: familiarity and applicability. The remainder of
the paper describes technical details of how to teach the standard
material of a CS theory course using non-decision problems. This brief
section tackles \emph{computability theory}, and the following section
tackles \emph{complexity theory}.

We say a computational problem is \emph{computable} if there exists a
program that solves it.  Note that computability is a generalization
of decidability, which applies only to decision problems.
Most undergraduate theory treatments use \emph{undecidability} as the central
concept to convey the profound idea that ``there are some problems
that computers can't solve.'' In the approach advocated by this paper,
the notion of \emph{uncomputability} replaces undecidability as the
central concept.  We can still use classical decision problems as
examples of uncomputable problems (e.g.\ the halting problem, or the
question of whether a given program computes a given function). But in
addition, we can discuss other interesting uncomputable problems that
are not decision problems.\footnote{Examples include the question of how many steps a program executes before terminating, or the length of a program's output. See~\cite{MacCormick2018PUP} for details.} In all cases, we retain the pedagogical
advantage of working in a framework that is perceived by the novice
audience as familiar and applicable.

\section{Elementary complexity theory for non-decision problems}
\label{sec:complexity-theory}

We now move from computability theory to complexity theory. Here, the
use of non-decision problems as the central focus requires some new
notation and terminology, compared to the traditional approach.  The
most fundamental classical complexity classes (\textsf{P},
\textsf{NP}, \textsf{Exp}) contain only decision problems, so we need
new notation for analogous classes that contain both decision and
non-decision problems.  Here, we will denote these new classes by
\textsf{Poly}, \textsf{NPoly}, and \textsf{Expo} respectively.
Formally, then, \textsf{Poly} is the set of computational problems for
which there exists a program (or Turing machine) that solves the
problem in polynomial time.  \textsf{NPoly} is the same but we allow
nondeterministic programs\footnote{The details of defining the
  outputs of nondeterministic programs that can produce more than a
  yes/no solution are interesting and important, but are omitted here
  for space reasons. See~\cite{MacCormick2018PUP} for details.}; and
\textsf{Expo} is the same but we allow exponential time.

The literature already has classes \textsf{FP} and \textsf{FNP}, which
are similar in spirit to \textsf{Poly} and \textsf{NPoly}
respectively---so we need good justification for introducing new
terminology into such well-trodden territory.  One reason is that
\textsf{FP} is sometimes defined in terms of function problems only
(which excludes problems with multiple solutions such as
\textsc{Factor} and \textsc{HamCycle}).  And even when \textsf{FP} is
defined in terms of general problems, it is usually done via the
polynomially bounded relations mentioned above, which is unnecessarily
complicated for the novice audience.  Therefore, we prefer to
introduce the new classes \textsf{Poly}, \textsf{NPoly}, and
\textsf{Expo}.

A comparison with our ten sample books is interesting here: five do not mention \textsf{FP}, \textsf{FNP}, or any
analogous complexity
class~\cite{davis1994computability,hopcroft2006introduction,lewis1997elements,linz2011introduction,Sipser2013};
three briefly cover one or both of \textsf{FP} and
\textsf{FNP}~\cite{Arora2009,Moore2011,rich2007automata}; and two give
more thorough analyses of \textsf{FP} and \textsf{FNP} or analogous
classes (Papadimitriou~\cite{Papadimitriou1994},
Goldreich~\cite{goldreich2010p}).  Papadimitriou's definitions of
\textsf{FP} and \textsf{FNP} are closest to the \textsf{Poly} and
\textsf{NPoly} defined here. However, they are not equivalent, because
of the restriction to polynomially balanced relations mentioned
above. Goldreich is the only one of the ten sample books to adopt
non-decision problems as a central concept. Goldreich defines the
interesting non-decision complexity classes \textsf{PF} (for
``polynomial time find'') and \textsf{PC} (for ``polynomial time
checkable''), but they are again not equivalent to \textsf{Poly} and
\textsf{NPoly}. Moreover, \textsf{PF} is not a subset of \textsf{PC}
(in contrast to the intuitive relationships
\textsf{P}$\subseteq$\textsf{NP},
\textsf{Poly}$\subseteq$\textsf{NPoly},
\textsf{FP}$\subseteq$\textsf{FNP}), making \textsf{PF} and
\textsf{PC} less than ideal for the novice audience. The overall
conclusion from analyzing the selection of textbooks is twofold: (i)
few textbooks discuss non-decision problems in any detail,
even via the well-established complexity classes \textsf{FP} and
\textsf{FNP}; and (ii) for technical reasons, the various non-decision
classes \textsf{FP}, \textsf{FNP}, \textsf{PF}, and \textsf{PC} are
unsuitable for the novice audience and we instead recommend
\textsf{Poly}, \textsf{NPoly}, and \textsf{Expo} as defined above.

Introducing novice audiences to complexity theory via the non-decision
classes \textsf{Poly}, \textsf{NPoly}, and \textsf{Expo} has a
striking advantage that is worth discussing further: the concrete
impact of polynomial time algorithms on cryptography is more
obvious. To see this, note that one way to crack the popular RSA
cryptosystem is to factor a large integer. So, if it turned out to
be true that \textsc{Factor}$\in$\textsf{Poly}, the extraordinary
consequences (namely, that RSA would be vulnerable to attack) are
immediately obvious to the novice audience. In contrast, the
consequences are unclear when the same concepts are taught in terms of
decision problems: we already know that
\textsc{FactorD}$\in$\textsf{P}, but that doesn't help to crack RSA
since we get only the existence and not the value of the factors.
There are ways of rephrasing factorization in terms of decision
problems,\footnote{Specifically, we can ask if there exists a factor
  within a given range. The search problem of finding a factor reduces
  to this decision problem, via binary search.}  but the extra
technical complexity of rephrasing the problem obscures the key point
for novice audiences.

What about more abstract problems such as \textsc{HamCycle} or
\textsc{Sat}? In these cases, the pedagogical advantages of
familiarity and applicability are compelling. To see why, first note
that both \textsc{HamCycle} and \textsc{Sat} can be reduced fairly
easily to their decision variants. Therefore, to an experienced
practitioner, it makes absolutely no difference whether we discuss
algorithms for \textsc{HamCycle} or \textsc{HamCycleD} (and similarly
for \textsc{Sat} vs \textsc{SatD}). But to the novice audience, the
act of finding a Hamilton cycle or satisfying assignment is much more
compelling than determining their existence. This distinction becomes
especially important when discussing nondeterminism, since
nondeterministic programs can easily and efficiently compute factors,
Hamilton cycles, and satisfying assignments. In fact, a beneficial and
enlightening homework assignment is for students to write
multithreaded programs that compute solutions to these problems in
nondeterministic polynomial time. As emphasized in
section~\ref{sec:empirical}, the fact that the outputs of such
programs are perceived as ``useful'' is an important factor in
achieving positive learning outcomes.


\section{The verifier-based definition of NPoly}
\label{sec:npoly-verifier-based}

It is well-known that \textsf{NP} has two equivalent definitions:
(i)~decision problems that can be decided by a polynomial time
nondeterministic program; and (ii)~decision problems whose positive
instances can be verified in polynomial time by a deterministic
program, when provided with a suitable \emph{certificate} (also known
as a \emph{witness}, or \emph{hint}).  Most modern textbooks cover
both definitions, usually introducing (ii) first and later proving the
equivalence to~(i).  Can we generalize these definitions to
non-decision problems in a way that appeals to the novice audience?
Yes we can, but with some caveats.

Definition~(i), based on nondeterminism, was discussed in the previous
section.
Here we focus on definition~(ii), based on verification. According to
this definition, \textsf{NPoly} is the class of computational problems
which have polynomial time verifiers. And \textsf{NP} is defined in a
precisely analogous way: it is the class of \emph{decision} problems
which have polynomial time verifiers. When we move to defining
\emph{verifier}, however, there are key differences in the decision
(\textsf{NP}) and non-decision (\textsf{NPoly})
scenarios. \conferenceVersion{Details of these differences can be
  found in the accompanying technical
  report~\cite{MacCormick:NonDecProbs:techreport:2017}. Here, we mention just one key
  aspect.} \techReportVersion{Before diving into these details, we
  give a high-level overview of one key aspect.} Recall that every
positive instance of an \textsf{NP} problem must be verifiable in
polynomial time when provided with some certificate $c$. Typically,
the certificate $c$ encodes a solution to the underlying non-decision
problem. For example, a \textsc{HamCycleD} instance can be verified by
providing a legitimate Hamilton cycle as the certificate. But
unfortunately, it turns out that certificates for \textsc{NP} problems
can also work in less intuitive ways. For example, it is possible to
define a \textsc{HamCycleD} verifier that uses only partial Hamilton
cycles as certificates.  This discrepancy between certificates and
underlying solutions is a source of confusion for the novice audience.

It turns out that if we work with non-decision problems, this
potentially confusing ``looseness'' in the definition of certificates
is eliminated. In essence (and please see \conferenceVersion{the
  technical report~\cite{MacCormick:NonDecProbs:techreport:2017} or
  textbook~\cite{MacCormick2018PUP}}
\techReportVersion{section~\ref{sec:verify-details}
  \nocite{MacCormick:NonDecProbs:techreport:2017}} for additional details), the
certificate $c$ is replaced by two separate strings: a solution $s$
and hint $h$. Both strings are given as input to the verifier. For
successful verifications of positive instances, the solution $s$
really is a solution for the given instance; the hint $h$ provides any
additional information needed for verification. The explicit roles of
$s$ and $h$ provide additional clarity to the novice audience.

To the best of our knowledge, no textbook other
than~\cite{MacCormick2018PUP} employs this formulation, which ensures
that verified solutions are meaningful while still permitting hints
for problems that need them.  This is one key difference between the
approach of this paper and that of Goldreich~\cite{goldreich2010p}.

\techReportVersion{

  \subsection{Details for verifying non-decision problems}
\label{sec:verify-details}

  For decision problems, the definition of verifier is brief and
  elegant but also surprisingly subtle. For example,
  quoting one of our sample books~\cite{Sipser2013}:
\begin{quote}
  A \emph{verifier} for a language $A$ is an algorithm $V$, where
\end{quote}
\vspace{-3mm}
\begin{equation} \label{eq:decision-verifier} \tag{$\star$}
  A = \{ w \,|\, V \text{ accepts } \langle w,c \rangle \text{ for some string } c \}.
\end{equation}
Of course, we are typically only interested in polynomial time
verifiers, where $V$ runs in polynomial time as a function of $|w|$.
In our sample of ten textbooks, all except
Goldreich~\cite{goldreich2010p} adopt concise, subtle definitions
similar to~(\ref{eq:decision-verifier}).  As with any such definition
in mathematics or computer science (a classic example is the
epsilon-delta definition of continuity in calculus), great care is
needed when presenting and explaining the definition to a novice
audience.  The formal definition should undoubtedly be presented, but
numerous examples demonstrating the full range of consequences are
equally important. In the particular case
of~(\ref{eq:decision-verifier}), key examples include:
\begin{enumerate}
\item \label{item:typical-V} The typical case in which, for a given
  positive instance $w$, the (polynomial time) algorithm $V$ accepts
  for only one special value of the certificate $c$, or perhaps a
  small number of special values.  For example, with
  \textsc{HamCycleD}, a given graph may have only a small number of
  Hamilton cycles.
\item The case of tractable problems (e.g.\ shortest path in a graph),
  in which $V$ can verify $w$ without the help of a
  certificate. Formally, this means $V$ accepts $\langle w,c \rangle$
  for \emph{all} $c$, even though the definition required acceptance
  only for \emph{some} $c$.
\item When $w$ is a negative instance (e.g.\ for \textsc{HamCycleD},
  $w$ is a graph with no Hamilton cycle), $V$ must reject $\langle w,c
  \rangle$ for \emph{all} $c$.  This makes good intuitive sense (e.g.\
  $V$ should never accept a certificate that claims to be a Hamilton
  cycle for a graph that doesn't actually have one).  The inversion of
  the quantifier (from \emph{some} to \emph{all}) is nevertheless a
  potential stumbling block for the novice audience.
\item \label{item:non-intuitive-V} Intuitively, the certificate $c$
  usually encodes a solution to the underlying non-decision problem.
  For concreteness, we illustrate this using \textsc{HamCycleD}. In
  this case, a typical implementation of $V$ might accept the
  certificate \str{a,b,c} whenever the tuple $(a,b,c)$ forms a
  Hamilton cycle in the instance $w$, and reject otherwise.  But there
  is no requirement for $V$ to behave in this way.  For example,
  another implementation of $V$ might permit the certificate to omit
  up to $k$ of the vertices at the end of the Hamilton cycle, for
  fixed $k$. By trying all the missing possibilities, $V$ can still
  correctly verify the suggested cycle in polynomial time.  Thus, a
  certain amount of ``looseness'' is deliberately built into the
  definition: the certificate need not explicitly contain a solution
  to the underlying problem, but we feel intuitively that a solution
  should be efficiently computable from the information in the
  certificate.
\end{enumerate}
Note that item~\ref{item:non-intuitive-V} is not a bug in the
definition~(\ref{eq:decision-verifier}).
Definition~(\ref{eq:decision-verifier}) works perfectly for the
complexity theory of decision problems. But
item~\ref{item:non-intuitive-V} does present problems when we
generalize the definition to non-decision problems, since we would
like to verify solutions explicitly (rather than verifying strings
from which a solution can be efficiently computed). The potential
discrepancy between certificates and solutions can be a source of
confusion for the novice audience, even if the material is presented
for decision problems only.

Items~\ref{item:typical-V}--\ref{item:non-intuitive-V} were presented
in detail as a kind of advance defense mechanism, because we will next
examine the generalized, non-decision definition of verifier. The
definition will initially seem more complex, and therefore apparently
unsuitable for the targeted novice audience. However, the key point
here is that any pedagogically sound discussion of the
apparently-simple definition~(\ref{eq:decision-verifier}) must also
include a discussion of
items~\ref{item:typical-V}--\ref{item:non-intuitive-V}.  When we move
to the non-decision variant of the definition,
items~\ref{item:typical-V}--\ref{item:non-intuitive-V} are largely
incorporated into the definition itself. However, the total ``degree
of difficulty'' of absorbing the definitions and their consequences is
similar. In fact, one might argue that in making the subtleties of the
definition explicit within the definition itself, we are facilitating
easier and deeper understanding by students.


\clearpage
So, here is the suggested definition for a verifier of non-decision
problems:
\begin{quote}
  Let $F$ be a computational problem. A \emph{verifier} for $F$ is a
  program $V(w,s,h)$ with the following properties:
  \begin{itemize}
  \item $V$ receives three string parameters: an instance~$w$, a
    proposed solution $s$, and a hint~$h$.
  \item $V$ halts on all inputs, returning either \str{yes} or
    \str{no}.
  \item \textbf{Every positive instance can be verified:} If
    $w$ is a positive instance of $F$, then $V(w,s,h)=\str{yes}$
    for \textbf{some} correct positive solution $s$ and \textbf{some}
    hint $h$.
  \item \textbf{Negative instances can never be verified:} If $w$ is a
    negative instance of $F$, then $V(w,s,h)=\str{no}$ for
    \textbf{all} values of $s$ and $h$.
  \item \textbf{Incorrect proposed solutions can never be verified:} If $s$ is
    not a correct solution (i.e.\ $s\notin F(w)$), then
    $V(w,s,h)=\str{no}$ for \textbf{all} $h$.
  \end{itemize} 
\end{quote}
In essence, the certificate $c$ in~(\ref{eq:decision-verifier}) has
been partitioned into a solution $s$ and hint $h$.  The hint $h$ plays
a similar role to $c$ in~(\ref{eq:decision-verifier}): $h$ may not be
needed at all, or it may provide only partial information (analogous
to the discussion above, in which the certificate omitted some
vertices from a cycle).  But we have more stringent conditions on $s$,
which ensure that it plays a meaningful role. With \textsc{HamCycle},
for example, $s$ is always a legitimate Hamilton cycle when $V$
accepts, and $V$ is guaranteed to reject any $s$ that is not a
legitimate Hamilton cycle.

Incidentally, $h$ is not required at all for \textsc{HamCycle},
because any solution is a complete certificate. For a related example
where $h$ plays a meaningful role, consider the problem
\textsc{HamCycleEdge}: the input is a graph and the solution set
consists of all edges that are members of some Hamilton cycle. Suppose
that the instance $w$ has a Hamilton cycle $(a,b,p,q,r)$. Then for $V$
to correctly and efficiently verify the solution $s=\str{a,b}$, it
would still need a hint indicating how to complete the Hamilton cycle:
$h=\str{p,q,r}$.

} 

\section{Reductions and NP-completeness}
\label{sec:reduct-np-compl}

We noted in the introduction that some of the theorems and proofs in
computational complexity theory are more elegant when expressed in
terms of decision problems, and the vast majority of literature takes
this approach.  Therefore, even if we choose for pedagogical reasons
to introduce elementary concepts in terms of non-decision problems, it
makes sense to eventually transition to the decision problem
viewpoint.  Over a four-year period, we experimented with making this
transition at various points in the course. For reasons discussed
below, we found the best choice is to make the transition when
students first encounter polynomial time reductions, and hence also
before introducing NP-completeness.

For the novice audience it seems preferable to focus on the simplest
type of reduction, which is variously known as a \emph{Karp reduction},
\emph{polynomial time mapping reduction}, or \emph{many-one
  reduction}. In focusing on Karp reductions, we are now in line with
eight of our ten sample textbooks (with the same two
outliers~\cite{goldreich2010p,Papadimitriou1994} as
previously). However, we still break slightly from the eight standard
treatments, because instead of defining reductions between
\emph{decision} problems, we define reductions from decision
computational problems to \emph{general} problems (i.e.\ \emph{to},
but not \emph{from}, general problems).  This difference again calls
for new terminology, and we use the term ``polyreduction'' for this.

Formally, a \emph{polyreduction} from the 
decision problem $D$ to the general computational problem $G$ is a map
$r:\Sigma^*\to\Sigma^*$, such that $r(w)$ is a positive instance of
$G$ if and only if $w$ is a positive instance of $D$. We also require
that $r$ is computable in polynomial time.  With this definition,
students can first examine trivial polyreductions such as
\textsc{HamCycleD}$\to$\textsc{HamCycle}, then move on to more
interesting ones such as
\textsc{DirectedHamCycleD}$\to$\textsc{UndirectedHamCycleD}.

With a little more effort, we can define polyreductions between
general problems $F$ and $G$. To do so, we need polynomial time maps
in both directions. First, $r:\Sigma^*\to\Sigma^*$ maps instances of
$F$ to instances of $G$. Then $r':\Sigma^*\to\Sigma^*$ maps solutions
of $G$ back to solutions of $F$.  The correctness condition is that,
given any program \texttt{G} that solves the problem $G$, we must have
$r'(\mathtt{G}(r(w)))\in F(w)$ for all $w$. (In words, the solution of
$G$ for $r(w)$ must map back to a solution of $F$ for $w$.) We have
experimented with teaching this concept to novice audiences and met
with a certain amount of success.  Goldreich also recommends this
approach~\cite{Goldreich2006}.  However, as already stated, on balance
we recommend avoiding this extra level of generality, instead
restricting polyreductions to be maps from decision problems to
general problems via the simpler definition in the previous paragraph.

In the same spirit, we might attempt to teach ``NPoly-completeness''
for general computational problems.  But in this case, our experience
has been that the benefits of using non-decision
problems are clearly outweighed by the disadvantages of dealing with
arbitrary solutions when stating and proving theorems about
NPoly-completeness.  Therefore, it seems preferable to stay solidly in
the traditional realm of decision problems when teaching
NP-completeness.  As a small bonus, the fact that our definition of
polyreduction allows decision problems to be reduced to \emph{general}
problems leads to an elegant definition of NP-hardness: a problem $G$
is NP-hard if some NP-complete problem $D$ polyreduces to
$G$. Holistic discussions of \textsf{P} versus \textsf{NP} (although
defined formally in terms of decision problems) also take on a more
practical tone when the majority of concepts earlier in the course
have been taught in terms of non-decision problems.

\section{Conclusion}
\label{sec:conclusion}

The paper advocated the use of non-decision problems in CS theory
courses. The paper consisted of two separate strands: (i)~an
explanation of \emph{why} non-decision problems are preferable, based
on a combination of empirical survey results and educational theory;
and (ii)~an explanation of \emph{how} to use non-decision problems in
the theory course, using reformulations of classical concepts.  In the
first strand (section~\ref{sec:empirical}), we demonstrated via an
empirical survey of CS majors that programs solving non-decision
problems are perceived as much more useful and applicable than
programs solving decision problems. Invoking the well-known
educational principle that learning outcomes are likely to be better
for a course that uses materials perceived as ``useful,'' we concluded
that non-decision problems should be used if possible. In the second
strand (sections~\ref{sec:practical-definition} and
\ref{sec:comp-non-decis}--\ref{sec:reduct-np-compl}), we suggested new
definitions, terminology, and notation that are designed to employ
non-decision problems and to maximize the accessibility of CS theory
concepts for the novice undergraduate audience. The approach has been
tested and refined over four years of teaching, culminating in a
recent textbook~\cite{MacCormick2018PUP}. Looking to the future, we hope
this approach will not only improve learning outcomes for
undergraduates receiving their first taste of computer science theory,
but will also lead to theory courses that are taught earlier in the
curriculum, to a wider range of undergraduates, at a wider range of
institutions.

\techReportVersion{
Finally, it should be mentioned that all definitions and technical
remarks in this paper would be obvious to experienced practitioners of
complexity theory, and no novelty for them is claimed. Nevertheless,
the suggestion of using these definitions and strategies to teach an
accessible theory course does appear to be novel.
}





\begin{thebibliography}{10}

\bibitem{Agrawal2004}
M.~Agrawal, N.~Kayal, and N.~Saxena.
\newblock {PRIMES} is in {P}.
\newblock {\em Annals of Mathematics}, 160(2):781--793, 2004.

\bibitem{Arora2009}
S.~Arora and B.~Barak.
\newblock {\em Computational Complexity: A Modern Approach}.
\newblock Cambridge University Press, 2009.

\bibitem{Chesnevar2003}
C.~I. Ches\~{n}evar, M.~L. Cobo, and W.~Yurcik.
\newblock Using theoretical computer simulators for formal languages and
  automata theory.
\newblock {\em SIGCSE Bulletin}, 35(2):33--37, June 2003.

\bibitem{Crescenzi2013}
P.~Crescenzi, E.~Enstr\"{o}m, and V.~Kann.
\newblock From theory to practice: {NP}-completeness for every {CS} student.
\newblock In {\em Proc. ITiCSE}, pages 16--21, 2013.

\bibitem{davis1994computability}
M.~Davis, R.~Sigal, and E.~J. Weyuker.
\newblock {\em Computability, Complexity, and Languages: Fundamentals of
  Theoretical Computer Science}.
\newblock Morgan Kaufmann, 2nd edition, 1994.

\bibitem{enstrom2010computer}
E.~Enstr{\"o}m and V.~Kann.
\newblock Computer lab work on theory.
\newblock In {\em Proc. ITiCSE}, pages 93--97, 2010.

\bibitem{fink2013creating}
L.~D. Fink.
\newblock {\em Creating significant learning experiences: An integrated
  approach to designing college courses}.
\newblock John Wiley \& Sons, 2nd edition, 2013.

\bibitem{fortnow2013golden}
L.~Fortnow.
\newblock {\em The golden ticket: P, NP, and the search for the impossible}.
\newblock Princeton University Press, 2013.

\bibitem{Goldreich2006}
O.~Goldreich.
\newblock {\em On Teaching the Basics of Complexity Theory}, pages 348--374.
\newblock Springer, 2006.
\newblock In {\em Theoretical Computer Science: Essays in Memory of Shimon
  Even}, ed. Goldreich, Rosenberg, Selman.

\bibitem{goldreich2010p}
O.~Goldreich.
\newblock {\em P, {NP}, and {NP}-Completeness: The Basics of Computational
  Complexity}.
\newblock Cambridge University Press, 2010.

\bibitem{hopcroft2006introduction}
J.~E. Hopcroft, R.~Motwani, and J.~D. Ullman.
\newblock {\em Introduction to Automata Theory, Languages, and Computation}.
\newblock Pearson, 3rd edition, 2006.

\bibitem{lewis1997elements}
H.~R. Lewis and C.~H. Papadimitriou.
\newblock {\em Elements of the Theory of Computation}.
\newblock Prentice Hall, 2nd edition, 1997.

\bibitem{linz2011introduction}
P.~Linz.
\newblock {\em An Introduction to Formal Languages and Automata}.
\newblock Jones \& Bartlett, 5th edition, 2011.

\bibitem{Lobo2006}
A.~F. Lobo and G.~R. Baliga.
\newblock {NP}-completeness for all computer science undergraduates: A novel
  project-based curriculum.
\newblock {\em J. Comput. Sci. Coll.}, 21(6):53--63, June 2006.

\bibitem{MacCormick:NonDecProbs:techreport:2017}
J.~MacCormick.
\newblock Strategies for basing the {CS} theory course on non-decision
  problems.
\newblock Technical report, Dickinson College, 2017.
\newblock Available as \url{http://arxiv.org/abs/xxxx.xxxx}.

\bibitem{MacCormick2018PUP}
J.~MacCormick.
\newblock {\em What Can Be Computed?: A Practical Guide to the Theory of
  Computation}.
\newblock Princeton University Press, 2018.

\bibitem{mandrioli1982teaching}
D.~Mandrioli.
\newblock On teaching theoretical foundations of computer science.
\newblock {\em ACM SIGACT News}, 14(3):36--53, 1982.
\newblock Part 2 appears in 14(4):58--69.

\bibitem{Moore2011}
C.~Moore and S.~Mertens.
\newblock {\em The Nature of Computation}.
\newblock Oxford University Press, 2011.

\bibitem{Papadimitriou1994}
C.~H. Papadimitriou.
\newblock {\em Computational Complexity}.
\newblock Addison Wesley, Massachussetts, 1994.

\bibitem{papadimitriou1997np}
C.~H. Papadimitriou.
\newblock {NP}-completeness: A retrospective.
\newblock In {\em International Colloquium on Automata, Languages, and
  Programming}, pages 2--6. Springer, 1997.

\bibitem{rich2007automata}
E.~Rich.
\newblock {\em Automata, Computability and Complexity: Theory and
  Applications}.
\newblock Pearson, 2007.

\bibitem{rodger2006jflap}
S.~H. Rodger and T.~W. Finley.
\newblock {\em JFLAP: an interactive formal languages and automata package}.
\newblock Jones \& Bartlett Learning, 2006.

\bibitem{rodger2009increasing}
S.~H. Rodger, E.~Wiebe, K.~M. Lee, C.~Morgan, K.~Omar, and J.~Su.
\newblock Increasing engagement in automata theory with {JFLAP}.
\newblock In {\em ACM SIGCSE Bulletin}, volume~41, pages 403--407. ACM, 2009.

\bibitem{Sipser2013}
M.~Sipser.
\newblock {\em Introduction to the Theory of Computation}.
\newblock Cengage, 3rd edition, 2013.

\end{thebibliography}

\end{document}